\documentclass[11pt,a4paper,twoside]{article} 
\usepackage{amsmath,amssymb,latexsym,epsfig,floatflt,subfigure,
  isolatin1}
\newcommand{\scr}{\scriptscriptstyle}

\newcommand{\muk}{\mu^{\scr +} K^0}
\newcommand{\epi}{e^{\scr +} \pi^0}
\newcommand{\et}{e^{\scr +} \eta}
\newcommand{\ek}{e^{\scr +} K^0}
\newcommand{\mpi}{\mu^{\scr +} \pi^0}
\newcommand{\mt}{\mu^{\scr +} \eta}
\newcommand{\ero}{e^{\scr +} \rho^0}
\newcommand{\eo}{e^{\scr +} \omega}

\newcommand{\mro}{\mu^{\scr +} \rho^0}
\newcommand{\mo}{\mu^{\scr +} \omega}
\newcommand{\nep}{\nu^{\scr C}_{\scr e} \pi^+}

\newcommand{\nmpi}{\nu^{\scr C}_{\scr \mu} \pi^+}
\newcommand{\nmk}{\nu^{\scr C}_{\scr \mu} K^+}
\newcommand{\nero}{\nu^{\scr C}_{\scr e} \rho^+}
\newcommand{\neks}{\nu^{\scr C}_{\scr e} K^{*+}}
\newcommand{\nmro}{\nu^{\scr C}_{\scr \mu} \rho^+}

\newcommand{\epin}{e^{\scr +} \pi^-}
\newcommand{\mpin}{\mu^{\scr +} \pi^-}
\newcommand{\eron}{e^{\scr +} \rho^-}
\newcommand{\mron}{\mu^{\scr +} \rho^-}
\newcommand{\nepn}{\nu^{\scr C}_{\scr e} \pi^0}
\newcommand{\nekn}{\nu^{\scr C}_{\scr e} K^0}
\newcommand{\nmpin}{\nu^{\scr C}_{\scr \mu} \pi^0}
\newcommand{\nmkn}{\nu^{\scr C}_{\scr \mu} K^0}
\newcommand{\neron}{\nu^{\scr C}_{\scr e} \rho^0}

\newcommand{\nmron}{\nu^{\scr C}_{\scr \mu} \rho^0}      

\newcommand{\neon}{\nu^{\scr C}_{\scr e} \omega}
\newcommand{\nmon}{\nu^{\scr C}_{\scr \mu} \omega}
\newcommand{\netn}{\nu^{\scr C}_{\scr e} \eta}
\newcommand{\nmtn}{\nu^{\scr C}_{\scr \mu} \eta}
\newcommand{\ntpn}{\nu^{\scr C}_{\scr \tau} \pi^0}
\newcommand{\ntkn}{\nu^{\scr C}_{\scr \tau} K^0}

\begin{document}
\def\NPB#1#2#3{Nucl. Phys. {\bf B} {\bf#1} (#2) #3}
\def\PLB#1#2#3{Phys. Lett. {\bf B} {\bf#1} (#2) #3}
\def\PRD#1#2#3{Phys. Rev. {\bf D} {\bf#1} (#2) #3}
\def\PRL#1#2#3{Phys. Rev. Lett. {\bf#1} (#2) #3}
\def\PRT#1#2#3{Phys. Rep. {\bf#1} C (#2) #3}
\def\ARAA#1#2#3{Ann. Rev. Astron. Astrophys. {\bf#1} (#2) #3}
\def\ARNP#1#2#3{Ann. Rev. Nucl. Part. Sci. {\bf#1} (#2) #3}
\def\MODA#1#2#3{Mod. Phys. Lett. {\bf A} {\bf#1} (#2) #3}
\def\NC#1#2#3{Nuovo Cim. {\bf#1} (#2) #3}
\def\ANPH#1#2#3{Ann. Phys. {\bf#1} (#2) #3}
\def\PTP#1#2#3{Prog. Th. Phys. {\bf#1} (#2) #3}

\begin{titlepage}
 
\thispagestyle{empty}
 
\begin{large}
\title{Gauge Mediated Proton Decay\\ in a Renormalizable
SUSY SO(10) with Realistic Mass Matrices}
\end{large}
 
\author{Yoav
  Achiman$^{a,b}$\footnote{e-mail:achiman@theorie.physik.uni-wuppertal.de}\quad
  \quad and \quad Marcus
  Richter$^b$\footnote{e-mail:richter@theorie.physik.uni-wuppertal.de}\\
  [1.5cm]
  {}$^a$School of Physics and Astronomy, Tel Aviv University, 69978 Tel Aviv,
  Israel\\
  {}$^b$Department of Physics, University of Wuppertal, D--42097 Wuppertal,
  Germany\\
  [1.5cm]} \date{ July 2001}
 
\maketitle
\setlength{\unitlength}{1cm}
\begin{picture}(5,1)(-12.5,-12)
\put(0,0){WUB 01-02}
\put(0,0.5){TAUP 2676-2001}
\end{picture}
\begin{abstract}
 Proton decay via d=5 operators is excluded by now not only in the framework
  of SUSY SU(5) but also its extensions like SUSY SO(10) are on the verge of
  being inconsistent with d=5 decays. 
  It is reasonable therefore to suppress, e.g. by a symmetry, the d=5 operators
  and to consider gauge boson induced d=6 decays. This is suggested in several
  recent papers in the framework of models with a lighter $ M_X$. 
  We discuss here 
  explicitly the fermionic sector of such a renormalizable SUSY SO(10) with 
  realistic mass matrices. We find that the recently ``observed''
  large leptonic mixing leads to an enhancement of the nucleon decay channels
  involving $\mu$'s and in particular the $\mu^+\pi^o$, $\mu^+\pi^-$ modes.  
\end{abstract}
\thispagestyle{empty}
 
\end{titlepage}

\newpage
\noindent
\parindent 0pt 
Nothing is known directly at present about the supersymmetric (SUSY) partners,
except for the experimental lower bounds on their masses. 
The main indication for (low energy) SUSY is the unification of the coupling
constant of the minimal SUSY Standard Model (MSSM) at a high scale
$\approx10^{16}$ GeV. However, this unification speaks even more for SUSY-Grand
Unified Theories (GUTs)~\cite{SG}, which predict also the observed partial
Yukawa unification (i.e. $m_\tau = m_b$ ) at those energies.  Yet proton 
decay, the other main prediction of GUTs, was not observed till 
now~\cite{pdecay}. Is this consistent with the expectation from SUSY GUTs?\\
 
It is generally assumed that the leading proton decay modes in SUSY-GUTs are
due to d=5 operators, as the d=4 contributions must be suppressed via a
symmetry (like R-parity) to avoid a much too fast decay. Those d=5 decay modes
are induced using the, not yet observed, SUSY particles and involve therefore
many unknown parameters~\cite{PD5}-\cite{alt}.  As a consequence the 
predicted rates are highly model dependent.  This large freedom does 
not save SUSY-SU(5) from being
practically ruled out if the needed threshold corrections are used to limit the
mass of the Higgsinos ($\tilde{H_t}$) which mediate proton
decay~\cite{N5}~\cite{Mu}.
Even extensions of SUSY SU(5) like SUSY SO(10) are on the verge of being
excluded by similar arguments~\cite{N10}\footnote{See however Refs.
 \cite{bpw} \cite{alt}.}.  
To save it one must push several parameters to their allowed
limit and in particular $M_{\tilde{H_t}}$ is required in some papers to be much
larger than $M_{GUT}$ and even larger than $\tilde{M}_{Planck}=2.4 \times
10^{18}$ GeV. This leads to very large couplings in the superpotential or the
corresponding non-renormalizable contributions.\\
 
In this letter we will assume, as was already put forward by several
people~\cite{D6}~\cite{Mu}, to suppress not only the d=4 but also the d=5 
contribution by a symmetry and consider the gauge mediated d=6 operators.
\begin{itemize}
\item{Those involve only known coupling constants and masses and 
therefore their predicted rates are far more reliable than the d=5 ones.} 
\item{They are the real test of the GUTs because d=5
induced proton decay is allowed in non-GUT theories as well.}  
\item{Although d=6 contributions are suppressed by $1/{M_X^4}$ they 
could yet be observable in the near future if the relevant gauge bosons 
mass $M_X$ is somewhat lower than $10^{16}$ GeV.}
\end{itemize}
 
This gauge-boson induced proton decay does not require non-renormalizable
contributions or particles with masses above the GUT scale. We will apply
therefore the ``renormalized see-saw mechanism'' using $\Phi_{\overline{126}}$
to give large masses to the right-handed (RH) Majorana neutrinos. This has many
advantages, in particular R-parity invariance, that is needed to avoid the
catastrophic d=4 contributions, is automatically obeyed in this case.  This
invariance leads also to a stable neutralino as a natural candidate for dark
matter.  On top of that, a way to lower the GUT scale in terms of a fully
renormalizable gauge theory was suggested recently by Aulakh, Bajc, Melfo,
Ra$\breve{s}$in and Senjanovi\'c~\cite{sen}.  They introduce an intermediate
Pati-Salam gauge group~\cite{ps} at a scale $M_I$.  Such an intermediate scale 
is needed anyhow to explain the fact that the masses of the heavy Majorana 
neutrinos required to give the light neutrinos masses consistent with the 
observed oscillations~\cite{nu}, are considerably smaller than the unification 
scale\footnote{This scale is needed also for the invisible-axion~\cite{ax} 
and the baryon asymmetry induced via the leptogenesis~\cite{bar}.}.
In this SUSY-SO(10) model, due to the absence of trilinear terms in the
superpotential, some particles acquire masses of $O(M_I^2/M_{GUT})$ via mixing.
Those particles affect the renormalization group equations in a way that lowers
the unification mass. The model can solve the doublet-triplet problem a la
Dimopoulos-Wilczek~\cite{DW} and can suppress naturally the d=5 operators i.e.
Higgsino mediated proton decay.
 
$M_X$ and/or the unification scale have smaller values also in other recent
models~\cite{D6} and in particular in those using large extra dimensions. As a
specially interesting example let us mention here the paper of Hall and
Nomura~\cite{hall} based on the model of Kawamura~\cite{kawamura}. 
The idea is
to use a five dimensional SU(5) GUT compactified to four dimensions on the
orbifold $S^1/(Z_2\times Z'_2)$.  This yields the MSSM with doublet-triplet
splitting and a vanishing proton decay from d=5 and d=4 operators by a $U(1)_R$
symmetry. The model gives a compactification scale ($M_c=M_X$) somewhat lower
than the four dimensional unification scale. This model is not a 4d GUT, but it
involves all the properties we use for the fermionic mass matrices and proton
decay.\\
 
When d=6 contributions are discussed in the literature one refers always to the
proton decay into $e^+\pi^0$, that is the dominant decay mode only when the
mixing is neglected.  This is in contrast with recent papers about d=5 proton
decay~\cite{N10}~\cite{bpw}, where the effects of realistic mass matrices are
explicitly considered.\\
 
We present here a model for the explicit realization of the fermionic sector
of a SUSY-SO(10) GUT with realistic fermionic mass
matrices. We will in particular show that the observed~\cite{nu} large 
leptonic mixing leads
to the enhancement of the branching ratios of the nucleon decay into muons with
respect to those with $e^+$ in the final state.
 
The observed large leptonic mixing will be coupled in our model with large 
quark mixing.
Note, that the CKM matrix gives the difference between the mixing angles 
of the up and down quarks. Therefore, only the {\em difference} between the 
lefthanded (LH) mixing angles of the quarks must be small while the 
righthanded (RH) rotations are unobservable in
the framework of the SM or the MSSM and can be large.
The nucleon decay in GUTs is one of the
few observables in which {\em all} mixing angles are involved\footnote{Partial
  information on the ``non-observable'' angles can be obtained by looking for
  RH currents, scalar leptoquark interactions, baryon asymmetry due to
  leptogenesis e.t.c.}.\\  

{\em The aim of our effective model is not to get as many as 
possible ``predictions'' for the known observables of the SM. It is to
calculate all the ``non-observable'' mixing angles in terms of the observables 
of the SM in order to predict the proton decay.}\\
 
The model uses a scheme developed in a series of papers~\cite{ag}~\cite{am}.
All mass matrices have a non-symmetric Fritzsch texture~\cite{hf}~\cite{NNI}
\begin{equation} 
{\bf M} \; = \; \left( \begin{array}{lll}
0 & A & 0 \\ B & 0 & C \\ 0 & D & E
\end{array} \right).
\end{equation}
This texture and the contributions of the Higgs representations are fixed by a
global $U_F(1)$ (or $Z_n$). We use only renormalizable contribution \`a l\'a
Harvey, Ramond and Reiss~\cite{hrr}.  This has some interesting advantages in
our opinion, as was mentioned before, compared to the method of Froggatt and
Nielsen~\cite{fn} that uses broken $U_F(1)$'s with non-renormalizable
contributions~\footnote{The method of Froggatt and Nielsen is useful to explain
  the hierarchy in the quark mass matrices but not for the neutrinos. This is
  because the matrix elements are fixed in this method only up to unknown
  $O(1)$ factors and the see-saw matrix is a product of three matrices.  The
  neutrino matrix elements are given in this case only up to corrections of
  $[O(1)]^3$ which can be large.}.
 
The SO(10) symmetry as well as $U_F(1)$ give relations between the different
entries of the mass matrices. We need one large VEV in ${\bf
  \Phi_{\overline{126}}}$ to give the RH Majorana neutrinos masses the
corresponding mass matrix must have the texture~\cite{ag}
\begin{equation} \label{mmans}
{\bf M_{\nu_R}} \; = \; \left( \begin{array}{lll}
0 & a & 0 \\ a & 0 & 0 \\ 0 & 0 & b
\end{array} \right).
\end{equation}
Because our main interest lies in the nucleon decay, CP violation is neglected
in this letter and the parameters are taken to be real, for simplicity~\footnote{Complex
parameters will be used in a forthcoming detailed paper~\cite{ar}}.

We use, as is discussed in~\cite{am}, one heavy ${\bf \Phi}_{\overline{\bf
    126}}$ to give the RH neutrinos a mass. To generate the light mass matrices
one ${\bf\Phi_{\overline{126}}}$ and pairs of ${\bf \Phi_{10}}$ and
${\bf\Phi_{\overline{120}}}$ are needed~\footnote{The model involves pairs like
  ${\overline{\bf126}}+ {\bf 126}$  e.t.c. to avoid breaking of the SUSY at
  high energies, but only part of those are relevant for the fermionic mass
  matrices.}. 
 Those Higgs representations involve several $SU_L(2)$ doublets. Note,
however that as
is usually done in SUSY GUT models~\cite{Raby}~\cite{ag} broken into the 
Minimal SUSY Standard Model(MSSM) at the TeV scale, only two combinations
of those doublets remain ``light''. All other combinations acquire heavy 
masses and therefore do not affect the gauge unification~\footnote{Some
  light VEVs are then induced ones.}. The explicit 
combinations are not important because the effective VEVs are anyhow free 
parameters in our model. For the ratio of the light VEVs we use in this letter
the commonly used value $\tan\beta = 5 $ ~\footnote{In Ref.~\cite{ar} we shall discuss a set of different values of $\tan\beta$.}. 
The gauge coupling becomes non-perturbative soon above the unification point
but we do not use physics above the GUT-scale.\\

The three fermionic families in ${\bf 16_i}$ i=1,2,3 and the
Higgs representations $\Phi_k$ transform under the global $U(1)_F$ as follows:
\begin{eqnarray}
{\bf 16}_j & \rightarrow & \exp(i \alpha_j \theta){\bf 16}_j \\
{\bf \Phi}_k & \rightarrow & \exp(i \beta_k \theta) {\bf \Phi}_k .
\end{eqnarray}
Invariance of the Yukawa coupling terms ${\bf 16}_i\Phi_k{\bf 16}_j$ under
$U(1)_F$ requires $ \alpha_i+\alpha_j=\beta_k$.  Hence, the most general
structure of the Yukawa matrices is:
\begin{eqnarray} \nonumber
{\bf Y}_{\bf \scr 10}^{\scr (1)} = \begin{pmatrix}
0 & x_1 & 0 \\ x_1 & 0 & 0 \\ 0 & 0 & \tilde x_1 \end{pmatrix} \; ; \quad
{\bf Y}_{\bf \scr 126}^{\scr (h)} = \begin{pmatrix}
0 & y_1 & 0 \\ y_1 & 0 & 0 \\ 0 & 0 & \tilde y_1 \end{pmatrix} \; ; \quad
{\bf Y}_{\bf \scr 120}^{\scr (1)} = \begin{pmatrix}
0 & z_1 & 0 \\ -z_1 & 0 & 0 \\ 0 & 0 & 0 \end{pmatrix} \; ; & & \\ \label{ycm}
{\bf Y}_{\bf \scr 10}^{\scr (2)} = \begin{pmatrix}
0 & 0 & 0 \\ 0 & 0 & x_2 \\ 0 & x_2 & 0 \end{pmatrix} \hspace{0.28cm} ; \quad
{\bf Y}_{\bf \scr 126}^{\scr (2)} = \begin{pmatrix}
0 & 0 & 0 \\ 0 & 0 & y_2 \\ 0 & y_2 & 0 \end{pmatrix} \hspace{0.22cm} ; \quad
{\bf Y}_{\bf \scr 120}^{\scr (2)} = \begin{pmatrix}
0 & 0 & 0 \\ 0 & 0 & z_2 \\ 0 & -z_2 & 0 \end{pmatrix} \hspace{0.25cm}. & &
\end{eqnarray}
These Yukawa matrices give explicit expressions for the ${\bf M_d}$,
${\bf M_u}$, ${\bf M_e}$ and
$ {\bf M}_\nu^{(Dir)}$ mass matrices, in terms of a set of 14 parameters
(combinations of the Yukawa couplings and the corresponding VEVs). 
These matrices are diagonalized and fitted to the the known masses
and mixing of the quarks and leptons as will be described in the following.\\ 

We start by taking diagonal mass matrices for the charged fermions at $M_Z$.
 Than the full two-loop renormalization group equations (RGEs) of the Minimal
Supersymmetric Standard Model (MSSM), with $\tan\beta=5$, are used to get 
the following masses at $M_{GUT}=2\times10^{16}$:

%
\begin{center}
\begin{tabular}{|c|c|c|}
\hline
$m_u(M_{\rm GUT})$ & $m_d(M_{\rm GUT})$ & $m_s(M_{\rm GUT})$ \\
\hline
1.04 MeV & 1.33 MeV & 26.5 MeV \\
\hline
\hline
$m_c(M_{\rm GUT})$ & $m_b(M_{\rm GUT})$ & $m_t(M_{\rm GUT})$ \\
\hline
302 MeV & 1 GeV & 129 GeV \\
\hline
\hline
$     m_e(M_{\rm GUT}) $ & $     m_\mu(M_{\rm GUT}) $ & 
$     m_\tau(M_{\rm GUT}) $ \\
\hline
0.32502032 MeV & 68.59813 MeV & 1171.4 MeV \\
\hline
\end{tabular}
\end{center}
%

In our model only 12 independent parameters are needed to specify charged
fermion matrices at the GUT scale, exactly the number of the underlying masses
and CKM mixing angles (the CKM matrix changes only slightly from $M_Z$ to
$M_{GUT}$). By fitting those parameters, {\em all} the LH and RH mixing angles
of the quarks and leptons are fixed\footnote{If only one parameter is
taken to be complex, one can use its phase to account for the observed
CP violation and again all mixing angles will be fixed~\cite{ar}.}. 
This procedure  involves a set of non-linear equations:
\begin{eqnarray} \nonumber
{\bf U}^{\dagger}_{\scr L} \, {\bf M}^{}_u \, {\bf U}^{}_{\scr R} & = &
{\bf M}^{\scr (D)}_u \; , \quad
{\bf D}^{\dagger}_{\scr L} \, {\bf M}^{}_d \, {\bf D}^{}_{\scr R} \ = \
{\bf M}^{\scr (D)}_d \; , \\ \label{soeq}
{\bf E}^{\dagger}_{\scr L} \, {\bf M}^{}_e \, {\bf E}^{}_{\scr R} \ & = &
{\bf M}^{\scr (D)}_e \; , \quad
{\bf U}^\dagger_{\scr L} {\bf D}^{}_{\scr L} \ = \ 
{\bf V}_{\scr \textrm{CKM}} \; .
\end{eqnarray}

We found five solutions to those equations all of which
have several large mixing angles~\cite{ar}. These fix the neutrino mass 
matrices ${\bf M}_{\nu}^{Dir}$ and ${\bf M_{\nu_R}}$ also up to
two parameters (and the overall scale $M_R$). Hence, the see-saw matrix
\begin{equation}
{\bf M}_\nu^{\scr \textrm{light}} \; \simeq \; -
{\bf M}^{\scr (\textrm{Dir})}_\nu \,
\big( {\bf M}^{\scr (\textrm{Maj})}_{\nu \scr{R}} \big)^{-1}
\big( {\bf M}^{\scr (\textrm{Dir})}_\nu \big)^T
\end{equation}
as well as the leptonic LH mixing 
${\bf U}_{MNS}= {\bf E}_{\scr L}^{\dagger} {\bf N}_\nu$ are also known. 
We varied than the two free parameters and looked for solutions that 
reproduced the neutrino data with LMA-MSW or SMA-MSW for the solar 
neutrinos~\cite{snu}.
The details will be given in a forthcoming paper~\cite{ar}.
The best fit is obtained for the a LMA-MSW solution with the 
following properties:

\begin{enumerate}  

\item The Quark LH and RH mixing angles at the GUT scale:
 
${\theta^u_L}_{12} = -0.077$,\quad ${\theta^u_L}_{23} = -1.48$,
\quad${\theta^u_L}_{13} = -4 \times 10^{-8}$.
 
${\theta^u_R}_{12} = -0.045$,\quad ${\theta^u_R}_{23} = -2.2\times
10^{-4}$,
\quad ${\theta^u_R}_{13} = -1.1 \times 10^{-3}$.
 
${\theta^d_L}_{12} = 0.15$,\quad ${\theta^d_L}_{23} = -1.44$,
\quad ${\theta^d_L}_{13} = 1 \times 10^{-5}$.
 
${\theta^d_R}_{12} = -0.33$,\quad ${\theta^d_R}_{23} = -3 \times 10^{-3}$,\quad
${\theta^d_R}_{13} = 6 \times 10^{-2}$.

\item The mixing angles of the Charged Leptons:

${\theta^\ell_L}_{12} = -1.17$,\quad ${\theta^\ell_L}_{23} = 1.44$,\quad
${\theta^\ell_L}_{13} = 0.0002 $.
  
${\theta^\ell_R}_{12} = 0.002$,\quad ${\theta^\ell_R}_{23} = -0.003$, \quad
${\theta^\ell_R}_{13} = 0.002 $.
 
\item The Neutrino masses :
 
$ M_R=5.2 \times 10^{13}$ GeV
 
$m_{\nu_e}= 1.88 \times 10^{-3}$ eV,\quad
$m_{\nu_\mu}= 5.89 \times  10^{-3}$ eV,\quad
$m_{\nu_\tau}=5.85 \times 10^{-2}$ eV.
 
\item The LH Leptonic (Neutrino) mixing angles:
 
$\theta^\nu_{12} = 0.55$,\quad
$\theta^\nu_{23} = 0.74$,\quad
$\theta^\nu_{31} = -0.0053$.
\end{enumerate}                           

Using these results one can calculate the proton and neutron decay branching
ratios. We use the method of Gavela et al.~\cite{gavela} as it was extended in
a series of papers~\cite{NSG}~\cite{am} to models with large
fermionic mixing (especially RH ones). In this work it is once more generalized
into a SUSY GUT.

The effective baryon number violating Lagrangian of SO(10) is
\begin{eqnarray} \nonumber
\mathcal{L}_{\scr \textrm{ef\/f}}
& = & A^{}_1 \, 
\big( \varepsilon_{\scr \alpha \beta \gamma} \bar u_{\scr L}^{\scr C
    \gamma}  \gamma^{\scr \mu} u_{\scr L}^{\scr \beta} \big)
\big( \bar e^{\scr +}_{\scr L} \gamma_{\scr \mu} d_{\scr L}^{\scr \alpha} \big)
\; + \; A^{}_2 \, 
\big( \varepsilon_{\scr \alpha \beta \gamma} \bar u_{\scr L}^{\scr C
    \gamma}  \gamma^{\scr \mu} u_{\scr L}^{\scr \beta} \big)
\big( \bar e^{\scr +}_{\scr R} \gamma_{\scr \mu} d_{\scr R}^{\scr \alpha} \big)
\\ \nonumber
& + & A^{}_3 \, 
\big( \varepsilon_{\scr \alpha \beta \gamma} \bar u_{\scr L}^{\scr C
    \gamma}  \gamma^{\scr \mu} u_{\scr L}^{\scr \beta} \big)
\big( \bar \mu^{\scr +}_{\scr L} \gamma_{\scr \mu} d_{\scr L}^{\scr
  \alpha} \big) 
\; + \; A^{}_4 \, 
\big( \varepsilon_{\scr \alpha \beta \gamma} \bar u_{\scr L}^{\scr C
    \gamma}  \gamma^{\scr \mu} u_{\scr L}^{\scr \beta} \big)
\big( \bar \mu^{\scr +}_{\scr R} \gamma_{\scr \mu} d_{\scr R}^{\scr \alpha} 
\big) \\ \nonumber
& + & A^{}_5 \, 
\big( \varepsilon_{\scr \alpha \beta \gamma} \bar u_{\scr L}^{\scr C
    \gamma}  \gamma^{\scr \mu} u_{\scr L}^{\scr \beta} \big)
\big( \bar e^{\scr +}_{\scr L} \gamma_{\scr \mu} s_{\scr L}^{\scr \alpha} \big)
\; + \; A^{}_6 \, 
\big( \varepsilon_{\scr \alpha \beta \gamma} \bar u_{\scr L}^{\scr C
    \gamma}  \gamma^{\scr \mu} u_{\scr L}^{\scr \beta} \big)
\big( \bar e^{\scr +}_{\scr R} \gamma_{\scr \mu} s_{\scr R}^{\scr \alpha} 
\big) \\ \nonumber
& + & A^{}_7 \, 
\big( \varepsilon_{\scr \alpha \beta \gamma} \bar u_{\scr L}^{\scr C
    \gamma}  \gamma^{\scr \mu} u_{\scr L}^{\scr \beta} \big)
\big( \bar \mu^{\scr +}_{\scr L} \gamma_{\scr \mu} 
s_{\scr L}^{\scr \alpha} \big) 
\; + \; A^{}_8 \, 
\big( \varepsilon_{\scr \alpha \beta \gamma} \bar u_{\scr L}^{\scr C
    \gamma}  \gamma^{\scr \mu} u_{\scr L}^{\scr \beta} \big)
\big( \bar \mu^{\scr +}_{\scr R} \gamma_{\scr \mu} s_{\scr R}^{\scr \alpha} 
\big) \\ \nonumber
& + & A^{}_9 \, 
\big( \varepsilon_{\scr \alpha \beta \gamma} 
\bar u_{\scr L}^{\scr C \gamma} \gamma^{\scr \mu} d_{\scr L}^{\scr \beta} 
\big)
\big( \bar \nu_{\scr e R}^{\scr C} \gamma_{\scr \mu} d_{\scr R}^{\scr
  \alpha}
\big)
\; + \; A^{}_{10} \, 
\big( \varepsilon_{\scr \alpha \beta \gamma} 
\bar u_{\scr L}^{\scr C \gamma} \gamma^{\scr \mu} d_{\scr L}^{\scr \beta} 
\big)
\big( \bar \nu_{\scr \mu R}^{\scr C} \gamma_{\scr \mu} d_{\scr
  R}^{\scr \alpha} \big)
\\ \nonumber
& + & A^{}_{11} \, 
\big( \varepsilon_{\scr \alpha \beta \gamma} 
\bar u_{\scr L}^{\scr C \gamma} \gamma^{\scr \mu} d_{\scr L}^{\scr \beta} 
\big)
\big( \bar \nu_{\scr e R}^{\scr C} \gamma_{\scr \mu} s_{\scr
  R}^{\scr \alpha} \big)
\; + \; A^{}_{12} \, 
\big( \varepsilon_{\scr \alpha \beta \gamma} 
\bar u_{\scr L}^{\scr C \gamma} \gamma^{\scr \mu} d_{\scr L}^{\scr \beta} 
\big)
\big( \bar \nu_{\scr \mu R}^{\scr C} \gamma_{\scr \mu} s_{\scr R}^{\scr
  \alpha} \big)
\\ \nonumber
& + & A^{}_{13} \, 
\big( \varepsilon_{\scr \alpha \beta \gamma} 
\bar u_{\scr L}^{\scr C \gamma} \gamma^{\scr \mu} s_{\scr L}^{\scr \beta} 
\big)
\big( \bar \nu_{\scr e R}^{\scr C} \gamma_{\scr \mu} d_{\scr R}^{\scr
  \alpha} \big)
\; + \; A^{}_{14} \, 
\big( \varepsilon_{\scr \alpha \beta \gamma} 
\bar u_{\scr L}^{\scr C \gamma} \gamma^{\scr \mu} s_{\scr L}^{\scr \beta} 
\big)
\big( \bar \nu_{\scr \mu R}^{\scr C} \gamma_{\scr \mu} d_{\scr
  R}^{\scr \alpha} \big)
\\ \nonumber
& + & A^{}_{15} \, 
\big( \varepsilon_{\scr \alpha \beta \gamma} 
\bar u_{\scr L}^{\scr C \gamma} \gamma^{\scr \mu} d_{\scr L}^{\scr \beta} 
\big)
\big( \bar \nu_{\scr \tau R}^{\scr C} \gamma_{\scr \mu} d_{\scr R}^{\scr
  \alpha} \big)
\; + \; A^{}_{16} \, 
\big( \varepsilon_{\scr \alpha \beta \gamma} 
\bar u_{\scr L}^{\scr C \gamma} \gamma^{\scr \mu} d_{\scr L}^{\scr \beta} 
\big)
\big( \bar \nu_{\scr \tau R}^{\scr C} \gamma_{\scr \mu} s_{\scr
  R}^{\scr \alpha} \big)
\\ \nonumber
& + & A^{}_{17} \, 
\big( \varepsilon_{\scr \alpha \beta \gamma} 
\bar u_{\scr L}^{\scr C \gamma} \gamma^{\scr \mu} s_{\scr L}^{\scr \beta} 
\big)
\big( \bar \nu_{\scr \tau R}^{\scr C} \gamma_{\scr \mu} d_{\scr R}^{\scr
  \alpha} \big) \\ \nonumber
& + & \; \textrm{(\, terms with two $s$ quarks \,)}
\\ \nonumber
& + & \; \textrm{(\, terms with $c$ ,$b$ and $t$ quarks \,)}
\\ \nonumber
& + & \; \textrm{(\, terms with $\bar \tau^{\scr +}_{\scr L,R}$ and 
$\bar \nu^{\scr C}_{\scr e,\mu,\tau L}$ \,)}
\\ \label{eldfnd}
& + & \textrm{h.c.}\\
\end{eqnarray}
where $A_i$ are functions of the mixing angles given in Ref.~\cite{am}.
 
The partial decay rate for a given process nucleon $\rightarrow$ 
meson + antilepton is expressed as follows~\cite{gavela}:
\begin{equation} \label{decform}
\Gamma^{}_j \; = \; \dfrac{1}{16\pi} \, m^2_{\scr \textrm{nucl}} \,
\rho_{\scr j} \, |S|^2 \, |\mathcal{A}_{}|^2 \,
\Big( |\mathcal{A}^{}_{\scr L}|^2 \sum_l |A^{}_{l} \mathcal{M}^{}_{l}|^2
+ |\mathcal{A}^{}_{\scr R}|^2 \sum_r |A^{}_{r} \mathcal{M}^{}_{r}|^2 
\Big) \; ,
\end{equation}
where $\mathcal{M}^{}_l$ and $\mathcal{M}^{}_r$ are the hadronic transition
matrix elements for the relevant decay process. $l$ and $r$ denote the
chirality of the corresponding antilepton~\cite{NSG}. $A^{}_{l}$ and $A^{}_{r}$ are the
relevant coefficients of the effective Lagrangian (\ref{eldfnd}).
$\mathcal{A}_{}$, $\mathcal{A}^{}_{\scr L}$ and $\mathcal{A}^{}_{\scr R}$ are
factors which result from the renormalization of the four fermion operators, 
 see Ref.~\cite{ar}.\\

Using all this we obtain the nucleon decay branching ratios for our best 
solution. These are presented in the following table compared with the
d=6 nucleon decays without mixing.\\

\centerline{
\hfill
\begin{tabular}{|l|r|r||l|r|r|}
\hline
proton  & \% & \% & neutron & \% & \% \\
decay channel  & no mixing  & LA-MSW & decay channel & no mixing & LA-MSW \\
\hline
\hline
$p \; \rightarrow \; \epi$  
& 33.6 & 17.5 & 
$n \; \rightarrow \; \epin$
& 62.86 & 32.5 \\
$p \; \rightarrow \; \mpi$  
& -- & 16.1 & 
$n \; \rightarrow \; \mpin$
& -- & 30.0 \\
$p \; \rightarrow \; \ek$   
& -- & 4.6 & 
$n \; \rightarrow \; \eron$
& 9.7 & 5.0 \\
$p \; \rightarrow \; \muk$  
& 5.8 & 2.7 & 
$n \; \rightarrow \; \mron$
& -- & 4.6 \\
$p \; \rightarrow \; \et$   
& 1.2 & 0.6 & 
$n \; \rightarrow \; \nepn$
& 15.1 & 9.2 \\
$p \; \rightarrow \; \mt$   
& -- & 0.6 & 
$n \; \rightarrow \; \nekn$
& -- & 2.6 \\
$p \; \rightarrow \; \ero$ 
& 5.1 & 2.7 & 
$n \; \rightarrow \; \netn$
& 0.6 & 0.3 \\
$p \; \rightarrow \; \mro$  
& -- & 2.5 & 
$n \; \rightarrow \; \nmpin$
& -- & 5.1 \\
$p \; \rightarrow \; \eo$   
& 16.9 & 8.8 & 
$n \; \rightarrow \; \nmkn$
& 1.7 & 0.0 \\
$p \; \rightarrow \; \mo$   
& -- & 8.1 &  
$n \; \rightarrow \; \nmtn$
& -- & 0.2 \\
$p \; \rightarrow \; \nep$  
& 32.3 & 19.7 &  
$n \; \rightarrow \; \neron$
& 2.3 & 1.4 \\
$p \; \rightarrow \; \nmpi$ 
& -- & 10.9 & 
$n \; \rightarrow \; \neon$
& 7.7 & 4.7 \\
$p \; \rightarrow \; \nmk$  
& 0.1 & 0.2 &  
$n \; \rightarrow \; \nmron$
& -- & 0.8 \\
$p \; \rightarrow \; \nero$ 
& 4.9 & 3.0 & 
$n \; \rightarrow \; \nmon$
& -- & 2.6 \\
$p \; \rightarrow \; \neks$ 
& -- & 0.1 & 
$n \; \rightarrow \; \ntpn$
& -- & 0.1 \\
$p \; \rightarrow \; \nmro$ 
& -- & 1.7 & 
$n \; \rightarrow \; \ntkn$
& -- & 0.7 \\
\hline
\end{tabular}
\hfill
}
\vskip 0.5cm

One sees clearly that the nucleon decay rates into muons are strongly enhanced.
Also decays into $e^+ K^0$ and $\nu K^0$ are not negligible.

The other solutions in the LMA-MSW case give similar results.  Since we use the
most general Yukawa matrices (in terms of the asymmetric Fritzsch texture), 
we expect our results to be generic in the large
mixing scenario. 

The absolute decay rates are much less reliable than the
branching ratios given above. They depend not only on the ``unknown'' value of
$M_X$, but also on the uncertain hadronic matrix elements~\cite{had}, the short
distance enhancement factors e.t.c.  
Using estimates for the absolute decay rates~\cite{D6}~\cite{Mu}, in terms
of the recent Lattice 
calculations of the JLQCD collaboration~\cite{JLQCD}, we obtain for our 
branching ratios the following result

$$
\frac{1}{\Gamma(p\rightarrow\mu^+\pi^o)} \approx \frac{1}{\Gamma(p\rightarrow
e^+\pi^o)}=16\times 10^{34}\times \Bigl(\frac{M_X}{10^{16}\hbox{GeV}}\Bigr)^4 
\Bigl(\frac{0.015\hbox{GeV}^3}{\alpha_H}\Bigr)^2 
{\hbox{ yrs}}.
$$

One can apply now the Super-Kamiokande bound~\cite{pdecay}

$$
\frac{1}{\Gamma(p\rightarrow e^+\pi^o)} > 4.4\times 10^{33}{\hbox{ yrs}}
{\hbox{ (90\% {\hbox{ CL}})}}
$$

to get a lower bound on effective $M_X$'s in the theory

$$
M_X > 0.7 \times 10^{16} \quad GeV\quad .
$$

To observe gauge mediated proton decay in the near future one needs surely
lighter X-bosons than $10^{16}$ Gev. This is a natural feature of several
recent models as indicated before.  
Super-Kamiokande will continue to look for the nucleon decay and expects to
reach in 8-years the limit of $1.4 \times 10^{34}$\quad years~\cite{sk}. 
Some of the new nucleon decay experiments
and especially ICARUS~\cite{ICARUS} and the half megaton UNO 
detector~\cite{UNO} are well suitable to look for decays into
$\mu$'s and $\nu$'s. Therefore, if the contributions of the d=5 operators to
the nucleon decay are really suppressed the search for the
$\mu^+\pi^o$, $\mu^+\pi^-$ modes should not be neglected.\\

The enhancement of the muon branching ratios is a unique feature of our
model because the decay mode $p \; \rightarrow \; e^+\pi^o$ is not negligible 
also in the $d=5$ induced decays~\cite{bpw}.
In view of the fact that this enhancement is the effect of the large 
observed leptonic mixing on  the $d=6$ nucleon decay, we suggest 
that {\em the observation of a considerable rate for the decay $p \; 
\rightarrow \;\mu^+\pi^o$ will be a clear indication for a gauge 
mediated proton decay}.\\

One can say in general, that the branching ratios of the nucleon decay can
teach us about the ``fundamental'' mass matrices as they depend on all mixing
angles. The present huge freedom in the mass matrices would then be strongly 
restricted and one could better understand the origin of the fermionic masses.

\end{document}